\documentclass[journal]{IEEEtran}
\usepackage{algorithm}
\usepackage{algorithmic}
\usepackage{multirow}
\usepackage{epsfig}
\usepackage{amssymb}
\usepackage{amsmath}
\usepackage{amsfonts}

\usepackage{subfig}             
\usepackage{array}
\usepackage{epsf}
\usepackage{epstopdf}
\usepackage{lineno,hyperref}
\modulolinenumbers[5]
\usepackage{graphicx}
\usepackage{threeparttable}
\usepackage[draft]{todonotes}   

\begin{document}

\title{The Role of Positive and Negative Citations in Scientific Evaluation}

\author{Xiaomei Bai, Ivan Lee, Zhaolong Ning, Amr Tolba, Feng Xia

\thanks{X. Bai, Z. Ning, F. Xia, are with the Key Laboratory for Ubiquitous Network and Service Software of Liaoning Province, School of Software, Dalian University of Technology, Dalian 116620, China.}
\thanks{X. Bai is also with Computing Center, Anshan Normal University, Anshan 114007, China.}
\thanks{I. Lee is with the School of Information Technology and Mathematical Sciences, University of South Australia, Australia.}
\thanks{A. Tolba is Riyadh Community College, King Saud University, Riyadh 11437, Saudi Arabia.}
\thanks{A. Tolba is also Mathematics and Computer Science Department, Faculty of Science, Menoufia University.}
\thanks{Corresponding Author: Z. Ning; Email: zhaolongning@dlut.edu.cn}

}
\markboth{IEEE Access}%
{Shell \MakeLowercase{\textit{et al.}}: Bare Demo of IEEEtran.cls for IEEE Journals}

\maketitle

\begin{abstract}
Quantifying the impact of scientific papers objectively is crucial for research output assessment, which subsequently affects institution and country rankings, research funding allocations, academic recruitment and national/international scientific priorities. While most of the assessment schemes based on publication citations may potentially be manipulated through negative citations, in this study, we explore Conflict of Interest (COI) relationships and discover negative citations and subsequently weaken the associated citation strength. PANDORA (Positive And Negative COI- Distinguished Objective Rank Algorithm) has been developed, which captures the positive and negative COI, together with the positive and negative suspected COI relationships. In order to alleviate the influence caused by negative COI relationship, collaboration times, collaboration time span, citation times and citation time span are employed to determine the citing strength; while for positive COI relationship, we regard it as normal citation relationship. Furthermore, we calculate the impact of scholarly papers by PageRank and HITS algorithms, based on a credit allocation algorithm which is utilized to assess the impact of institutions fairly and objectively. Experiments are conducted on the publication dataset from American Physical Society (APS) dataset, and the results demonstrate that our method significantly outperforms the current solutions in Recommendation Intensity of list R at top-K and Spearman's rank correlation coefficient at top-K.
\end{abstract}

%
%
\section{Introduction}
\IEEEPARstart{W}{ith} the development of scholarly big data, quantifying research output plays an important role in ranking institutions and countries, allocating research grants, making hiring decisions and planning scientific priorities~\cite{van2013maze,Xia2017Big,bai2017overview}. For instance, it is observed that the institutional appraisal system has been consistently evolving to address the various needs in scientific impact evaluation~\cite{aragon2012measure,bornmann2014ranking,bornmann2014effect}. While self-citations are meant for reflecting the research progress or knowledge diffusion which is a standard and acceptable procedure~\cite{G2017Assessing}, the act could, however, be abused to artificially inflate research impact~\cite{Bartneck2011Detecting} or introduce excessive self-advertisement~\cite{Seglen1992The}. In this paper, the scope of self-citations is extended to cover co-authors citations (note: co-author citations are not limited to co-authors of the citing paper, but co-authors in all papers over a pre-defined time frame, such as over the past three years), as well as colleague citations (i.e. co-workers belong to the same institute). The act of the extended scope of self-citation, which is referred as COI citations in this paper, may either be necessary or negative as discussed above. Prior study shows that self-citation accounts for $36\%$ under a three-year citation window~\cite{Aksnes2003A}, and this figure is expected higher under COI as it's an extended scope of self-citation. Unfortunately, there¡¯s lack of universal ground truth to differentiate legitimate citations among all COI citations. Thus, there's a need to develop a mechanism to model COI citations, and associate different weights to different type of citations. It should be noted that, legitimate self-citations should be treated as regular citations with standard weights (usually 1); whereas negative citations should be given a lower weight (i.e. this weight may be less than 1) to reduce their inflated impact.

Current approaches of institution appraisal fall in two primary categories: full counting and fractional counting. The full counting based methods assume that one scholarly paper contributes equally to all authors' institutions, so the impact of the paper could be counted multiple times~\cite{Vinkler2010The}. The fractional counting based methods consider the rate indicators of a selected list of top journals and the best papers or focus only on highly-cited papers~\cite{bornmann2014mapping}.
Fractional counting may also suffer from potential distortion in institutional impact evaluation. This is because diversity of negative citations~\cite{yao2014ranking,Bai2016Identifying} is neglected. A feasible solution to the problem is to design a better assessment metric, which can identify the negative citations, assesses the impact of academic papers objectively and allocates fair shares to contributing institutions.

The evaluation techniques for the impact of scientific outputs have evolved dramatically in recent years, from citation-based indicators~\cite{garfield2006history} to rank-based metrics~\cite{yao2014ranking,sayyadi2009futurerank,wang2013ranking}. A nonlinear PageRank algorithm was proposed to improve the effectiveness of reference ranking~\cite{yao2014ranking}. FutureRank was presented to rank scientific articles by citation, authors and time~\cite{sayyadi2009futurerank}. Based on FutureRank, Wang et al.~\cite{wang2013ranking} proposed a CAJTRank to fairly evaluate scientific articles by considering citation, authors, journals and time information. Due to the rapid development of impact metrics, some researchers appealed to exploiting a measured method for metrics~\cite{Wilsdon2015We,nature2013beware}. Unfortunately, most existing metrics regard all citations with equal weights, and it is likely that these metrics do not properly reflect accurate impact of authors, journals and institutions. Therefore, there is a crucial need for an improved metric to fairly assess academic papers.

There are two technical challenges for fair appraisal of academic affiliations: (1) identifying negative citations; and (2) distinguishing the citation strength, and allocating the impact of scientific paper to the affiliation(s) of each author. Dissecting the citation relationship is a feasible way for solving the first problem, and we define positive/negative COI relationships and positive/negative suspected COI relationships as follows:

Positive/negative COI relationships: for two papers with existing citing relationship, if the authors of the two papers are ever co-authors, and if there are one or more papers citing the two papers at the same time, the citation behavior is viewed as positive COI relationships. Otherwise, if no independent papers (papers from different authors) co-cite the two papers, the citation is considered as a negative COI relationship.

Positive/negative suspected COI relationships: similar to the definition of Positive/negative COI, which leverages citation behavior among coauthors, suspected COI leverages the citation behaviors for authors from the same affiliation. Likewise, if there are independent papers (papers without authors from the same institute) recognizing the correlation of papers from the same affiliation, the citation is viewed as an positive suspected COI, otherwise the citation is considered as a negative suspected COI.


We propose PANDORA to model both the COI and suspected COI relationships for fairly measuring the impact of papers. We leverage the credit allocation algorithm~\cite{shen2014collective} to capture coauthor¡¯s contributions of a paper, and adjusted weighting is applied to distribute the impact of a publication to affiliation(s) of each author.

\begin{figure}[htbp]
  \centering
  \includegraphics[width=0.45\textwidth]{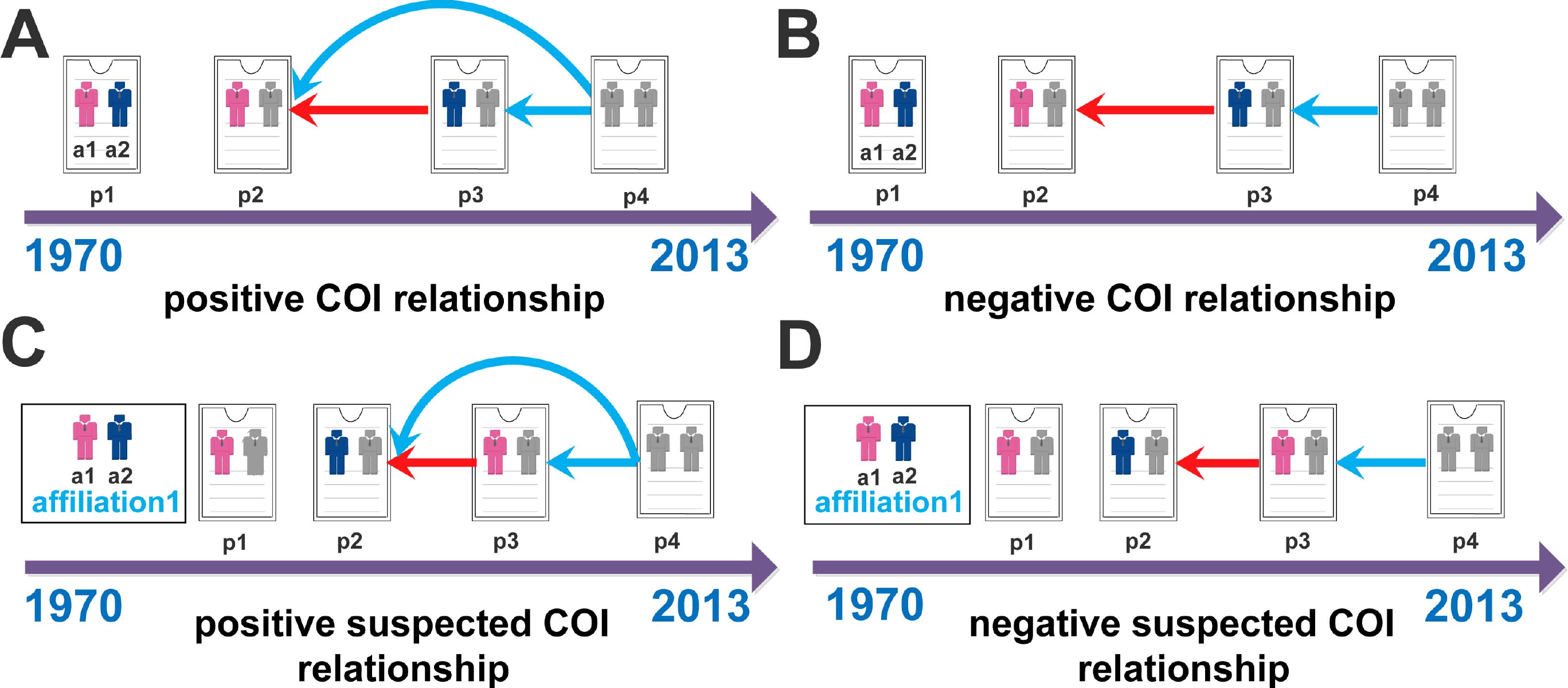}
  \caption{Illustrative example of different COI relationships between authors.}
  \label{Figure1}
\end{figure}

In this paper, we differentiate positive and negative COI relationships, positive and negative suspected COI relationships according to citation patterns, as illustrated in Fig.~\ref{Figure1}.
The measurements of negative COI and negative suspected COI relationships between researchers employ the following four factors: collaboration times, collaboration time span, citation times and citation time span. It should be noted that if the authors of paper $P_i$ citing paper $P_j$ is a positive COI, the citation is regarded normal, and the citation strength is set as 1. Extensive experiments are conducted on two subsets of American Physical Society (APS), i.e. PRC and PRE. The results demonstrate that our method outperforms the existing approaches in the Recommendation Intensity at top-K and Spearman¡¯s rank correlation coefficient at top-K.

(Note: where $P_i$ and $A_i$ are the list of papers and authors, respectively. The red line and blue line denote the citing relationship. The figure illustrates four cases: (A) Before $P_i$ cites $P_j$, the authors of $P_i$ and $P_j$ have co-authored one or multiple publications, and $P_i$ and $P_j$ are co-cited by other publications. Just like $P_3$ cites $P_2$, author $A_1$ and author $A_2$ ever collaborated $P_1$, and $P_2$ and $P_3$ were co-cited by $P_4$, $(A_1,A_2)$ makes up a positive COI author pair. (B) Compared with (A), $P_i$ and $P_j$ are not co-cited by other publications, just like $P_2$ and $P_3$ have not been cited by other papers simultaneously, $(A_1,A_2)$ composes a negative COI author pair. (C) Before $P_i$ cites $P_j$, between authors of $P_i$ and $P_j$ have not collaborated each other, but $A_i$ and $A_j$ belong to the same affiliation, and $P_i$ and $P_j$ are co-cited by other papers, just like the relationship of $P_3$ and $P_2$, $(A_1,A_2)$ is a positive suspect COI author pair. (D) Compared with (C), $P_i$ and $P_j$ are not co-cited by other publications, as $P_3$ and $P_2$, $(A_1,A_2)$ composes a negative suspect COI author pair.)

\section{Design of PANDORA}
\subsection{Dataset}
Our experiments are conducted on the APS dataset, which contains 71,287 publications published in two different subsets of APS: Physical Review C and Physical Review E (http://publish.aps.org/datasets), between 1970 and 2013 (43 years). Each record in the dataset includes the paper¡¯s title, DOI, author(s), date of publication, affiliation(s) of authors and publisher. A list of citations is provided by an independent dataset within the APS dataset.

\subsection{Impact of a scholarly paper}
\begin{figure}[htbp]
  \centering
  \includegraphics[width=0.45\textwidth]{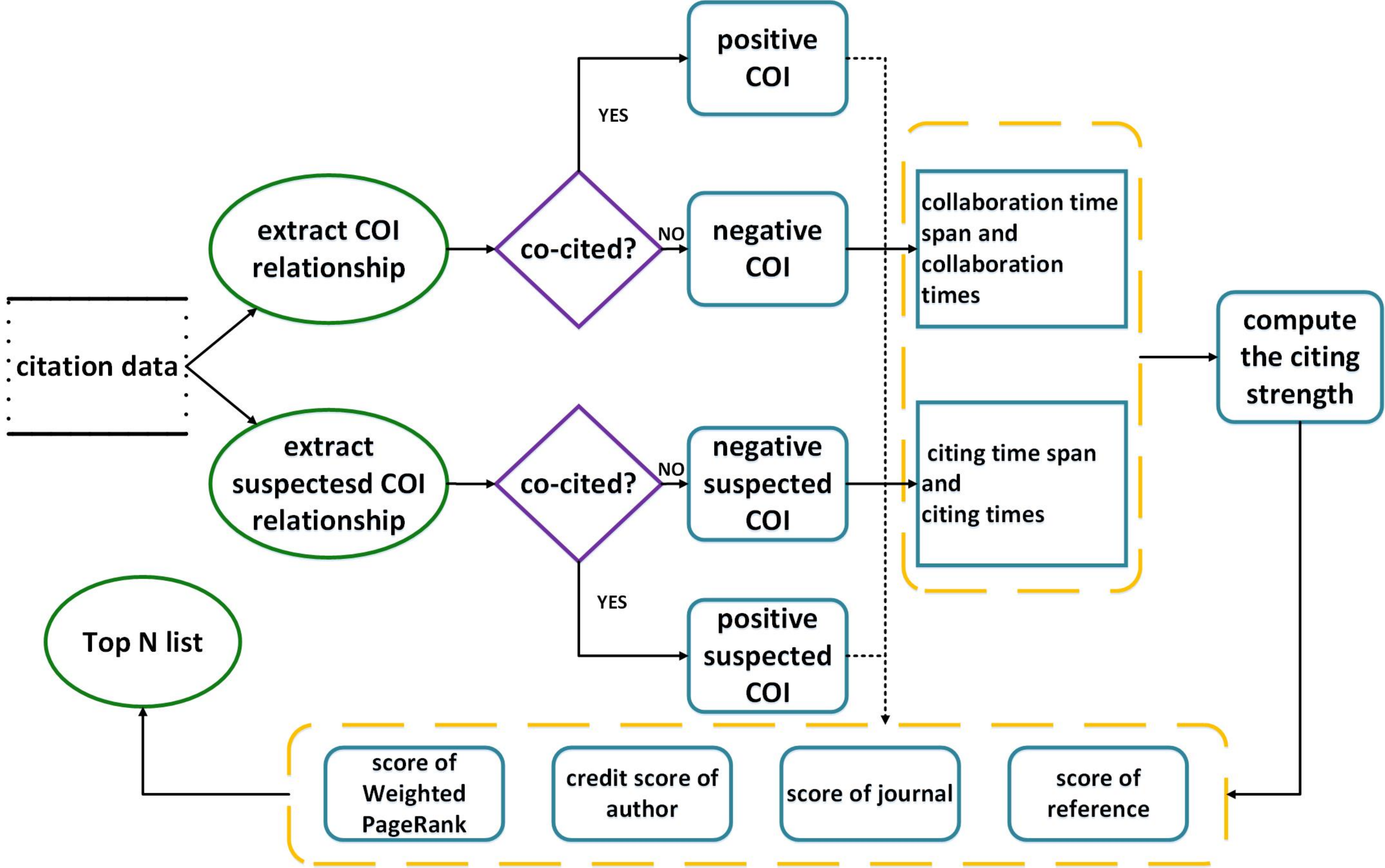}
  \caption{The structure of PANDORA. (Note: Ranking the impact of publications contains three steps: (1) Extract positive COI, negative COI, positive suspected COI and negative suspected COI from APS dataset; (2) Calculate citing strength according to the relationship of step (1); (3) List top N papers by the weighted PageRank algorithm and HITS algorithm.)}
  \label{Figure2}
\end{figure}

The structure of PANDORA is illustrated in Fig.~\ref{Figure2}. PANDORA quantifies publication impact by differentiating positive COI, negative COI, positive suspected COI and negative suspected COI between the citing and cited publications. Furthermore, PANDORA computes the authoritative score of each paper based on CAJTRank~\cite{wang2013ranking}. CAJTRank was a graph-based ranking method and used PageRank~\cite{Page1998The} algorithm and HITS~\cite{Kleinberg1998Authoritative} algorithm simultaneously to rank scientific papers by relying on the four factors: citations, authors, journals (conferences) and publication time. PANDORA adopts the weighted PageRank to fairly evaluate the impact of scholarly papers, and it also leverages the credit allocation algorithm~\cite{shen2014collective} to allocate the impact of a paper to its signed authors. In this paper, we consider citations, authors, journals and time factors, mainly because these factors can influence the impact of papers.

\subsection{Identification of positive COI}
If the authors of a citing paper and a cited paper have been co-authors of one or more papers, COI relationships exist. Although the COI citations may distort the appraisal of research impact, some of these citations are positive and necessary. Consider a scenario where the authors of the citing paper $P_i$ and the cited paper $P_j$ have published papers together. If $P_i$ and $P_j$ are co-cited by the papers from different authors, we consider the citation of $P_i$ and $P_j$ is reasonable, and we refer the citation as a positive COI, with its citation strength $W_{i,j}^{P-PCOI}$ set as 1.

\subsection{Identification of negative COI}
If COI relationships exist between the authors of a citing paper $P_i$ and a cited paper $P_j$, and no paper cites $P_i$ and $P_j$ simultaneously, we define cases like this as negative COI. The citing strength is quantified by the correlation strength between the authors, and we introduce two factors: collaboration times, and collaboration time span, for adjusting the weight of citation strength. The negative COI strength of co-authors in PANDORA is defined as follows:
\begin{equation}\label{eq:e1}
  W_{x,y}^{A-NCOI}=\frac{N_{x,y}^{Co-author}}{\Delta T_{c}}
\end{equation}
where $W_{x,y}^{A-NCOI}$ is the negative COI strength of co-authors between the $x$th author and the $y$th author. $N_{x,y}^{Co-author}$ indicates the cumulative number of papers coauthored by the $x$th author and the $y$th author. $N_{x,y}^{Co-author}=|S_x \cap S_y|$, where $S_x$ is a set of papers published by the $x$th author, and $S_y$ is a set of papers published by the $y$th author. $T_{x,y}^N$ is the last year of the $x$th author citing the $y$th author. $T_{x,y}^1$ is the initial year of the $x$th author starts citing the $y$th author. $\Delta T_{c}=T_{x,y}^N-T_{x,y}^1+1$  indicates the number of year between the first and the last collaborations of authors $A_x$ and $A_y$.

The relationship strength of each two publications based on the co-author's negative COI is calculated by:
\begin{equation}\label{eq:e2}
  W_{i,j}^{P-NCOI}=\sum_{i}^{I}\sum_{j}^{J}(\frac{N_{i,j}^{Co-paper}}{\Delta T_{c}})
\end{equation}
where $W_{i,j}^{P-NCOI}$ is the negative COI strength of the $i$th paper and the $j$th paper. It is the sum of negative COI between each two authors of citing paper and cited paper. $N_{i,j}^{Co-paper}$ indicates the cumulative number of papers coauthored between each two signed authors in the $i$th paper and the $j$th paper. $N_{i,j}^{Co-paper}=|S_i \cap S_j|$, $S_i$ is the set of all papers with signed authors of the $i$th paper. $S_j$ is the all papers set of signed authors of the $j$th paper. The citation strength of each two publications is defined as follows:
\begin{equation}\label{eq:e3}
  W_{i,j}^{P-NCite}=e^{-\rho(T^{Current}-T^{Cite}+1)W_{i,j}^{P-NCOI}}
\end{equation}
where an exponentially decaying formula is proposed to calculate the citation strength between paper $P_i$ and paper $P_j$. We consider the citing strength ranges from 0 to 1. As scholars usually intend to cite recent work, if a paper fails to attract citations in early years after being disseminated, the chance of attracting more citations over time will be limited. $W_{i,j}^{P-NCite}$ is the citation strength of the $i$th paper citing the $j$th paper. $W_{i,j}^{P-NCite}$ is defined within a range between 0 and 1 in our work, instead of assuming the citation strength as 1 (i.e. without regarding the COI relationships) in previous works. The value of citation strength is reasonable with the range between 0 and 1, and the formula favors current citation. $\rho$ is a constant value illustrating predefined decay parameter~\cite{wang2013ranking}. $T^{Current}$ stands for the current time, $T^{Cite}$ is the time of paper $P_i$ citing paper $P_j$ , and $T^{Current}-T^{Cite}+1$ represents the time duration of paper $P_i$ citing paper $P_j$.

\subsection{Identification of positive suspected COI}
Given paper $P_i$ cites paper $P_j$, although the authors between the two papers have not collaborated in producing any joint paper, the authors of $P_i$ and $P_j$ belong to the same affiliation, we consider that suspected COI relationship exists between them. If the two publications are co-cited by one or more publications (i.e. demonstrated relevance), we regard paper $P_i$ citing paper $P_j$ as positive suspected COI, and the citing strength $W_{i,j}^{P-PSCOI}$ is set as 1.

\subsection{Identification of negative suspected COI}
Given paper $P_i$ cites paper $P_j$, the authors between the two papers have not co-authored any paper, and the authors of $P_i$ and $P_j$ belong to the same affiliation, however the two publications are not co-cited by other publications (i.e. without demonstrated relevance), we regard paper $P_i$ citing paper $P_j$ as negative suspected COI. The citing strength is quantified by introducing the two factors: citing times and citing time span. The strength of negative suspected COI relationship of each two authors is defined as follows:
\begin{equation}\label{eq:e4}
  W_{x,y}^{A-NSCOI}=\frac{N_{x,y}^{Cite}}{\Delta T_{s}}
\end{equation}
where $W_{x,y}^{A-NSCOI}$ is the negative suspected COI strength of co-authors of the $x$th author and the $y$th author. $N_{x,y}^{Cite}$ is the cumulative number of papers of the $x$th author citing the $y$th author. $\Delta T_{s}=T_{x,y}^N-T_{x,y}^1+1$ indicates the number of years between the first and the last citing of authors $A_x$ and $A_y$. The strength of suspected COI relationship of each two articles is calculated by:
\begin{equation}\label{eq:e5}
  W_{i,j}^{P-NSCOI}=\sum_{i}^{I}\sum_{j}^{J}(\frac{N_{i,j}^{Cite}}{\Delta T_{s}})
\end{equation}
where $W_{i,j}^{P-NSCOI}$ is the negative suspected COI strength of the $i$th paper and the $j$th paper. It is a summation of all the negative suspected COI between authors of paper $P_i$ and paper $P_j$. $N_{i,j}^{Cite}$ is the cumulative citing number between the authors of the $i$th paper and the $j$th paper. The strength of citation relationship of each two articles is defined as:
\begin{equation}\label{eq:e6}
  W_{i,j}^{P-NSCite}=e^{-\rho(T^{Current}-T^{Cite}+1)W_{i,j}^{P-NSCOI}}
\end{equation}

\subsection{The authority score of a paper}
In CAJTRank, all citing weights are set as 1, and the abnormal citations are ignored, which leads to the potential issues in article impact evaluation. PANDORA is proposed to address these issues to improve objective and accurate impact assessment of research articles. The core idea of PANDORA is that the prestige of a publication is quantified by the scores of its weighted-PageRank, authors, journal published and references, namely the impact of a publication. Compared to CAJTRank algorithm, PANDORA method constructs a weighted-PageRank to capture the authority score of each publication in citation networks. While the initial score of each paper is set as $1/N$, $N$ indicates the total numbers of scholarly papers in the experiment. Meanwhile, PANDORA also considers the authority scores of each author, journal and references, which are calculated by HITS algorithm. Particularly, CAJTRank algorithm assumed all co-authors' contributions are equal to a paper. The CAJTRank algorithm neglects a fact that contributions of different authors of a paper are never equal. In order to resolve this problem, credit allocation is introduced in order to reasonably distribute the influential score of different authors of individual paper~\cite{shen2014collective}. Concrete process of computing the prestige score of each publication is demonstrated as follows:

The weighted PageRank score of paper $P_i$ is computed by
\begin{equation}\label{eq:e7}
    weightedPageRank(P_i)=\sum_{P_j\in IN(P_i)}\frac{W_{j,i}}{|OUT(P_j)|}S(P_j)
\end{equation}
where $IN(P_i)$ indicates all the papers of linking to paper $P_i$, $|OUT(P_j)|$ is the total number of publication $P_j$ linking out. $W_{j,i}$ illustrates the citation strength of paper $P_j$ citing paper $P_i$. $S(P_j)$ is the original score of paper $P_j$ before iteration is updated.

The authors' authority scores of single publication are determined by the influence score of each author, while the prestige score of individual author is related to the impact scores of his/her published papers, and can be calculated by the HITS algorithm. When computing the hub score of each author, credit allocation algorithm is used to distribute the proportion of impact of single paper to different authors.

Specifically, the authors' prestige scores of each paper, $Author(P_i)$, is calculated by
\begin{equation}\label{eq:e8}
\begin{split}
    Author(P_i)=&\frac{1}{T(A)}\cdot \sum_{A_{j}\in Neighbor(P_{i})}\\
     &\frac{\sum_{P_{k}\in Neighbor(A_{j})}CreditShare(P_{k},A_{j})\cdot S(P_{k})}{|Neighbor(A_{j})|}
\end{split}
\end{equation}
where $T(A)$ indicates a summation of all authors' hub scores in the experimental data. $Neighbor(P_i)$ denotes the co-authors' list of paper $P_i$, $Neighbor(A_j)$ shows all the papers published by author $A_j$, $CreditShare(P_k, A_j)$ illustrates the proportion of publication $P_k$ impact score by different authors of $P_k$. $S(P_k)$ is the prestige score of paper $P_k$, $|Neighbor(A_j)|$ is the number of publications of author $A_j$.

Likewise, the prestige score of a journal paper is computed by the HITS algorithm. Specific formula is shown as follows:
\begin{equation}\label{eq:e9}
\begin{split}
    Journal(P_i)=&\frac{1}{T(J)}\cdot\\
    &\sum_{J_{j}\in Neighbor(P_{i})}\frac{\sum_{P_{k}\in Neighbor(J_{j})}S(P_{k})}{|Neighbor(J_{j})|}
\end{split}
\end{equation}
where $Journal(P_i)$ demonstrates the prestige score of publication $P_i$ transmitted from its published journal, and the hub score of each journal includes the prestige scores of all the papers published in the journal. $T(J)$ is the sum of all journals' hub scores, $Neighbor(P_i)$ is the journal published by paper $P_i$, and each paper belongs to one journal. $Neighbor(J_j)$ is the set of papers published by journal $J_j$, $|Neighbor(J_j)|$ is the total number of publications in $Neighbor(J_j)$.

The reference scores of each publication are also computed by the HITS algorithm:
\begin{equation}\label{eq:e10}
\begin{split}
    Reference(P_i)=&\frac{1}{T(P)}\cdot\\
    &\sum_{P_{j}\in Neighbor(P_{i})}\frac{\sum_{P_{k}\in Neighbor(P_{j})}S(P_{k})}{|Neighbor(P_{j})|}
\end{split}
\end{equation}
where $Reference(P_i)$ shows the publication score $P_i$ collected from the authority scores of its references. $T(P)$ indicates the total scores transmitted from all the hub papers. $Neighbor(P_j)$ contains all the publications that $P_j$ links to. $|Neighbor(P_j)|$ is the sum of publication references $P_j$.

The authority score of each article contains the four components: PageRank score of the paper, scores of authors, journal and references of this work. The authority score is defined as the weighted sum of these components plus a normalize term:
\begin{equation}\label{eq:e11}
\begin{split}
    S(P_i)=&\alpha\cdot weightedPageRank(P_i)+\beta\cdot Author(P_i)\\
    &+\gamma\cdot Journal(P_i)+\delta\cdot Reference(P_i)\\
    &+(1-\alpha-\beta-\gamma-\delta)\cdot \frac{1}{N}
\end{split}
\end{equation}

In our experiment, if the score gap between the present and previous prestige of each paper is less than 0.0001, we consider the presented PANDORA iterative algorithm converges. $S(P_i)$ indicates the authority score of publication $P_i$. $\alpha$, $\beta$, $\gamma$ and $\delta$ are constant parameters, ranging from 0 to 1. The probability of random jump is set as 0.15 experimentally; meanwhile, the summation of $\alpha+\beta+\gamma+\delta$ is set as 0.85 for obtaining good experimental results. Currently, the typical approaches for parameter estimation include simple linear regression, multivariate linear regression and support vector machine regression~\cite{purwins2014regression}. Based on our experimental characteristic, multivariate linear regression is employed to estimate the parameters of PANDORA, CAJTRank and FutureRank, due to the characters of different factors and linear evaluation. We set the sum of all parameters as 0.85 in the three algorithms, and we find that a parameter captures relative high value, and other parameters receive relative low values by multivariate linear regress. We then estimate three groups of optimal parameters to compare the accuracy of RI and Spearman¡¯s rank correlation coefficients of the above mentioned three algorithms. More details can be found in reference~\cite{jiang2012towards}.

\subsection{Impact of institution and country}
In order to assess the impact of each institution objectively, the computing process is divided into the two parts in this section. The first part is to objectively allocate the impact of each publication to its co-authors by PANDORA. If a publication is only signed by an author, the prestige score of the paper belongs to the single author and this situation is relatively popular in decades ago. With the development of Internet, the collaboration among multiple authors also starts to prevail. Correspondingly, fair distribution of publication impact to multiple authors becomes an important practice to evaluate the impact of different scholars and their institutions.

To fairly distribute the impact of a paper, credit allocation algorithm is leveraged to resolve the coauthors' contributions to a paper. In this algorithm, the contributive proportion of each author to a paper is determined by the total credit from all of his/her co-cited papers. The second part is to calculate the impact of different institutions. The impact of a scholar is defined as follows:
\begin{equation}\label{eq:e12}
    I_A=\sum _{i}R\cdot S(P_{i})
\end{equation}
where $I_A$ refers to the set of individual scholar's impact captured from his/her publications. $R$ refers to the proportion of credit allocation to different authors. $S(P_i)$ indicates the prestige score of a scholarly paper.

The institutional impact is determined by the impact of all scholars' publications in this institution:
\begin{equation}\label{eq:e13}
    I_I=\sum _{i}I_{A,i}
\end{equation}
where $I_I$ indicates the set of institutional impact, containing the impact of all scholars in this institution. $I_{A,i}$ denotes a scholar's impact. For the authors with multiple affiliations, we consider the first affiliation as the primary institution.

Furthermore, the impact of a country is determined by the impact of publications of all institutions in this country. Given a country with $i$ institutions, the impact of a country is defined as follows:
\begin{equation}\label{eq:e14}
    I_C=\sum _{i}I_{I,i}
\end{equation}
where $I_C$ represents the set of country impact. $I_{I,i}$ denotes the institutional impact.

\section{Results}
\subsection{Measuring the impact of papers}
In order to objectively evaluate the impact of institutions, we first explore the COI relationships in citation networks. 71,287 papers with 755,902 author pairs are analyzed in our experiment. The author pairs are divided into the four categories: positive COI, negative COI, positive suspected COI and negative suspected COI. It is interesting to note that the author pairs of positive COI relationships and negative COI relationships are 59,388 pairs and 48,377 pairs respectively. There are 25,803 papers with positive COI, and 24,294 papers about with negative COI. We also find that there are 3,196 author pairs exist suspected COI relationships, in which author pairs of positive suspected COI relationships and negative suspected COI relationships are 1,645 and 1,551 respectively.
%

We conduct the experiments on two subsets of APS to compare PANDORA with FutureRank~\cite{sayyadi2009futurerank} and CAJTRank~\cite{wang2013ranking}. The experiments use two metrics: $RI$~\cite{jiang2012towards} and Spearman¡¯s rank correlation coefficient~\cite{myers2010research}. Due to the lack of ground truth for ranking papers, previous research adopted the future PageRank score~\cite{sayyadi2009futurerank} or the future citations~\cite{wang2013ranking} as the ground truth. In our study, citations without COI are used as the ground truth instead. The reason is that both future PageRank scores and future citations contain negative citations. Therefore, the ground truth may be biased, and it likely results in the inaccurate appraisal of the impact of academic publications.

The concept of $RI$ is described as follows: let $R$ indicate the list of top-K returned papers of a ranking method and $L_1$ represent the list of ground truth. For any scholarly paper $P_i$ in $R$ with the ranked order $ro$, the $RI$ of $P_i$ at $k$ is defined as:
\begin{equation}\label{eq:e15}
RI(P_{i})@k=
\left\{
\begin{array}{rr}
1+(k-ro)/k, P_{i}\in L_{1} \\
0, P_{i}\notin L_{1}
\end{array}
\right.
\end{equation}

The above formula implies that for any article $P_i$ of the top-K ground truth list, if it is ranked higher, $RI$ of the article $P_i$ will be higher. Correspondingly, we are able to gain the $RI$ of the list $R$ at $k$. According to the $RI$ of each article, the $RI$ of the list $R$ at $k$ could be defined as follows:
\begin{equation}\label{eq:e16}
RI(P_{i})@k=\sum_{P_{i}\in R}RI(P_{i})@k
\end{equation}

\begin{figure}[htbp]
  \centering
  \includegraphics[width=0.45\textwidth]{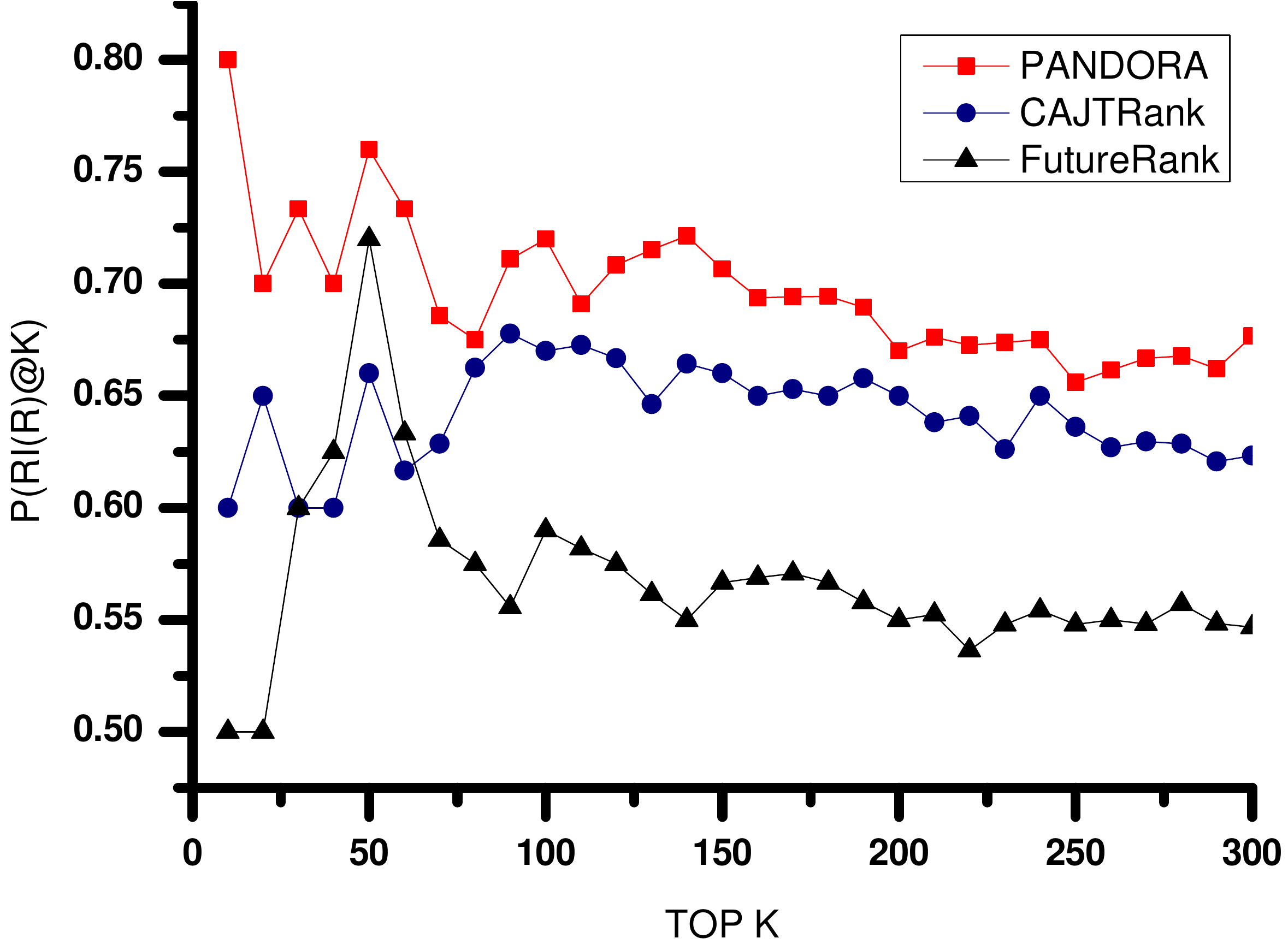}
  \caption{Comparing the probabilities of Recommendation Intensity for PANDORA, FutureRank and CAJTRank.}
  \label{Figure3}
\end{figure}

To better understand the impact that PageRank, authors, journal, citations and time factors have on the prestige scores of each article, a multivariate linear regression is implemented to estimate the parameters of PANDORA, FutureRank and CAJTRank, to optimize the ranking results in terms of $RI$ performance. Fig.~\ref{Figure3} shows different $RI$ accuracy rates of the three algorithms on top-K papers, and $K$ ranges from 10 to 300. The $RI$ accurate rates of CAJTRank and FutureRank are between 0.6 and 0.677, and between 0.5 and 0.633, respectively. In comparison, the $RI$ accurate rates of PANDORA are between 0.656 and 0.8. As shown in Fig.~\ref{Figure3}, PANDORA consistently outperforms CAJTRank and FutureRank in terms of $RI$ for all $K$ values.

Moreover, the Spearman¡¯s rank correlation coefficient is utilized to assess the similarity of these algorithms.
\begin{equation}\label{eq:e17}
\rho =\frac{\sum_{i}(R_{1}(P_{i})-\overline{R}_{1})(R_{2}(P_{i})-\overline{R}_{2})}
{\sqrt{\sum_{i}(R_{1}(P_{i})-\overline{R}_{1})^{2}\sum_{i}(R_{2}(P_{i})-\overline{R}_{2})^{2}}}
\end{equation}
where $R_1(P_i)$ and $R_2(P_i)$ indicate the position of publication $P_i$ in the ground truth rank list and the corresponding algorithm rank list, respectively. $\overline{R}_{1}$ and $\overline{R}_{2}$ are the average rank positions of all publications in the two ranks lists, respectively.

\begin{figure}[htbp]
  \centering
  \includegraphics[width=0.45\textwidth]{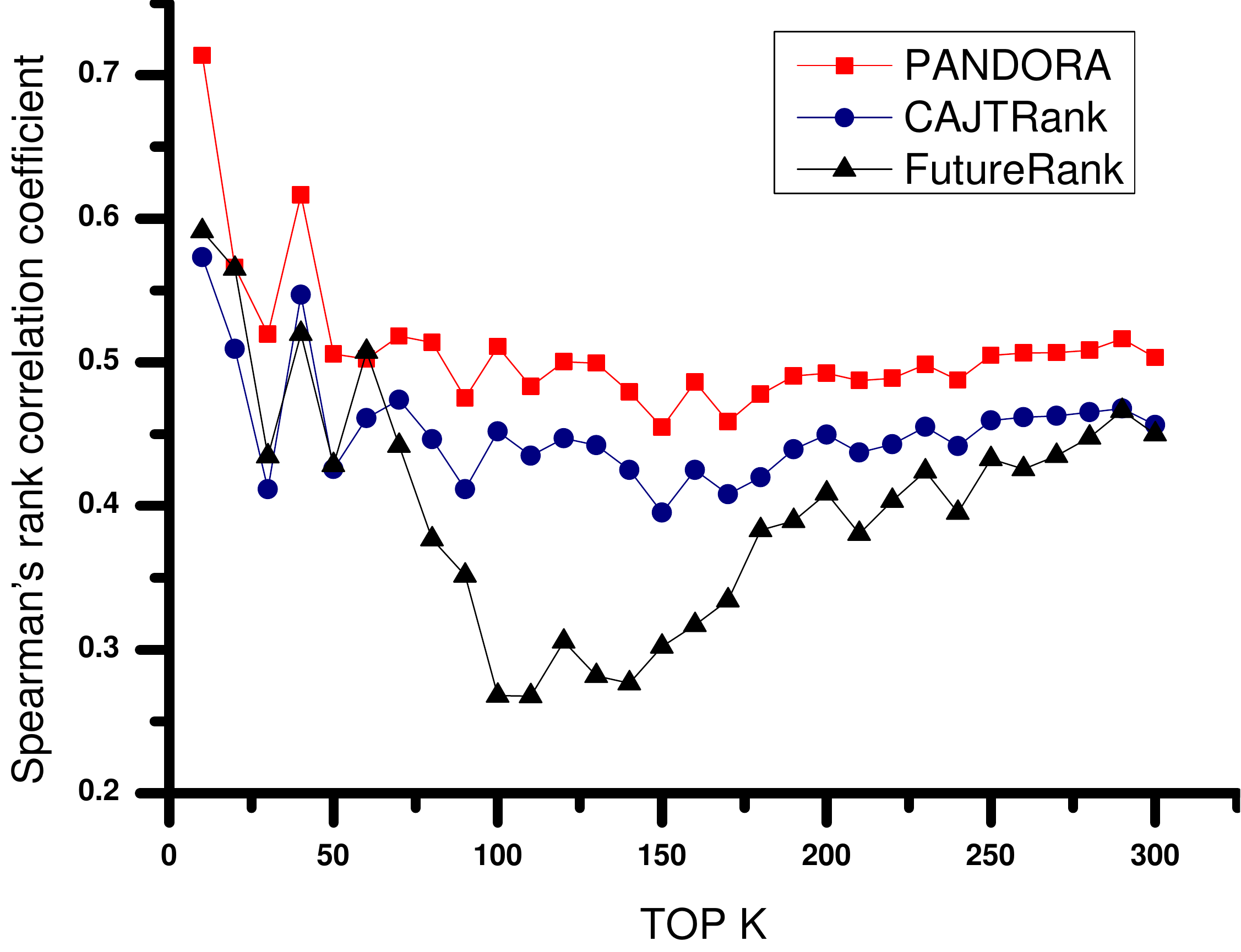}
  \caption{Plots of the Spearman¡¯s rank correlation coefficient against top-K papers with PANDORA, FutureRank and CAJTRank algorithms.}
  \label{Figure4}
\end{figure}

In Fig.~\ref{Figure4}, we observe that Spearman's rank correlation coefficients of PANDORA and CAJTRank are at around 0.5 and 0.42 respectively, while the coefficient changes from 0.267 to 0.519 for FutureRank. The best result could be obtained by considering all kinds of information: PageRank, authors, journal, references and time factors. PANDORA outperforms the other state-of-art methods on two subsets of APS.

By comparing $RI$ and Spearman's rank correlation coefficients of CAJTRank and FutureRank, we find that journal factor is beneficial to assess the impact of papers. At the same time, by comparing PANDORA and CAJTRank, the preceding results indicate that by the using COI relationships, weighted PageRank could improve the performance of PANDORA. Next, we take dynamic evolutionary nature of citation networks, time information and COI relationships of authors into consideration, and jointly use them to determine the citing strength, which have important effect on generating a better $RI$ and Spearman's rank correlation coefficient. Moreover, PANDORA's author credit allocation scheme helps generating a fair and objective ranking list.

\subsection{Quantifying the impact of institutions}
In order to capture the existence of COI citations at the institution-level and at the country-level, several characteristics are investigated for institution $m$ and country $n$: the institution size ($A_m$), represents the total number of authors from institution $m$ who published at least one publication. $P_m$ shows the number of publications published from institution $m$. $COIC_m$ is the cumulative COI citations derived from all publications in institution $m$. The country size ($A_n$), denotes the total number of authors from country $n$ who published at least a publication. $P_n$ represents the number of publications published from country $n$; $COIC_n$ shows the cumulative COI citations derived from all publications in country $n$.

\begin{figure*}[htbp]
  \centering
  \includegraphics[width=1\textwidth]{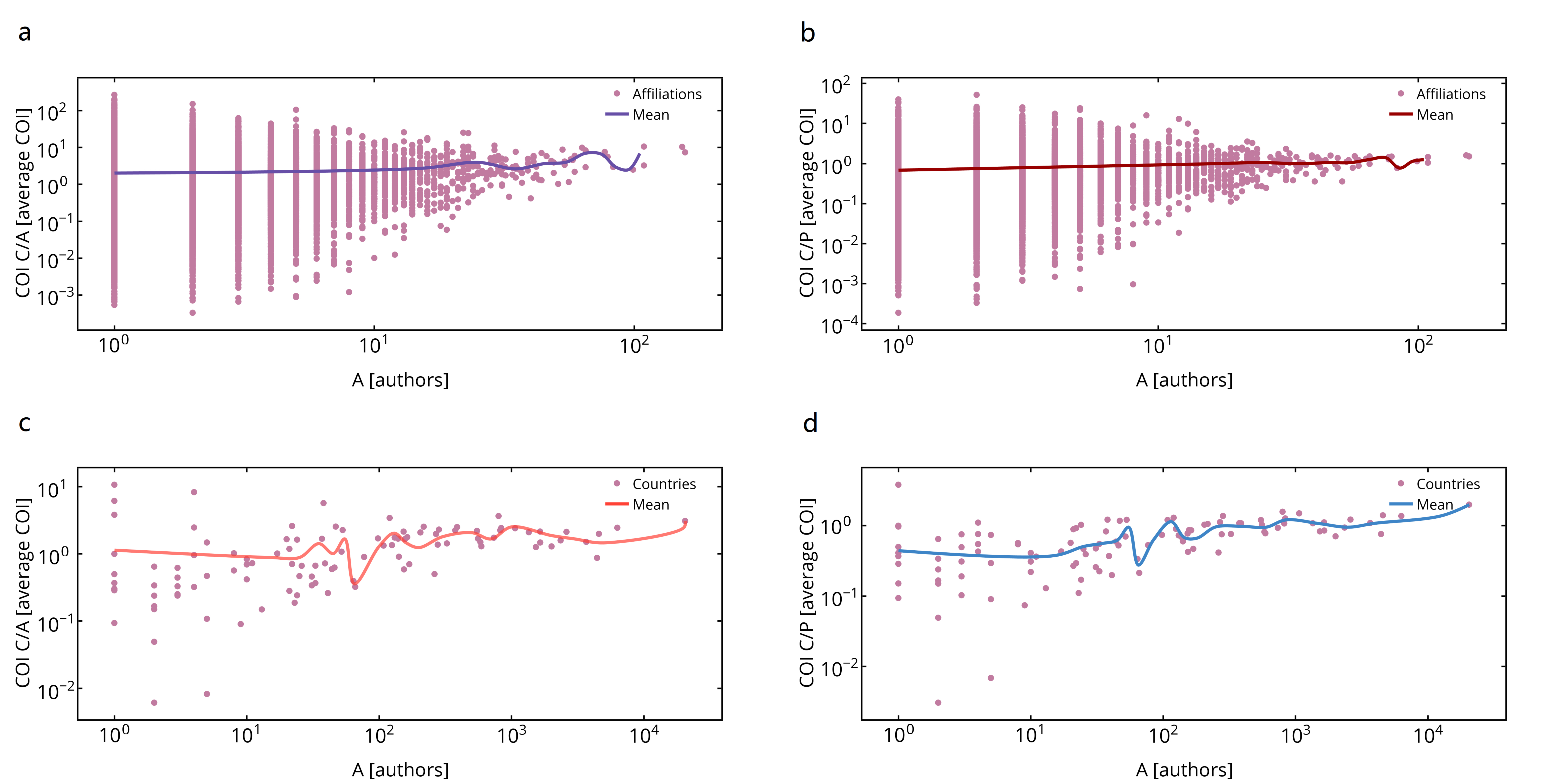}
  \caption{Quantitative patterns of COI citations for institutions and countries. (Note: (a) The institution size vs. the average COI citations is shown, indicating most institution sizes are very small, and a few have large scholars. (b) The correlation between institution size and the average publications COI citations shows institution size has little influence on the $COIC/P$. (c) The correlation between Country size and each scholar's average COI citations indicates that large countries have more authors with COI citations. (d) The relationship between country size and the average COI citations of each publication illustrates that large countries have more papers with COI citations.)}
  \label{Figure5}
\end{figure*}

Figs.~\ref{Figure5}a and \ref{Figure5}b show the correlation between the institution size $A_m$ and both the average COI citations $COIC/A_m$ and the average publications COI citations $COIC/P$. Fig.~\ref{Figure5}a indicates that most institutions have a small number of scholars publishing on APS journals. It could be observed that COI citations exist in institution with different scales, and the larger the institution is, the more COI citations are likely to be. Fig.~\ref{Figure5}b shows the correlation between institution size $A_m$ and the average COI citations of each paper $COIC/P_m$. According to Fig.~\ref{Figure5}b, we can observe that COI citations of each paper range from $10^{-4}-10^{2}$, and the average COI citations of each paper range from 0.529 to 1.022.
Figs.~\ref{Figure5}c and \ref{Figure5}d show the correlation between the country size $A_n$ and both the average COI citations $COIC/A_n$ and the average publications COI citations $COIC/P_n$. Fig.~\ref{Figure5}c indicates that most countries have a small number of scholars in the APS publications, except for a few countries. For example, American scholars are the highest contributors to PRC and PRE, with 20,468 scholars. We also find that the country size $A_n$ positively correlates with COI citations, which indicates that large countries are likely to result in COI citations according to the analysis of each author's COI citations. The potential cause of this phenomenon is that large institutions or large countries have more international and internal collaborations. According to Fig.~\ref{Figure5}c, we also observe that when country size falls in between ranks 60 and 100, $COIC/A_n$ presents a sudden fall. To obtain a better understanding of the decrease, we investigate these countries and their adjacent countries in the following three aspects: the number of papers, the number of co-authors and COI citations. We observe an interesting phenomenon: small COI citations exist in these countries, and the number of co-authors (about 3 authors) of each paper is less than their adjacent countries (around 4 authors). This may be the reason why impact of these countries is relatively small compared with other countries. Fig.~\ref{Figure5}d shows the correlation between the country size $A_n$ and the average COI citations of each paper $COIC/P_n$. It indicates that each publication contains more COI citations in large countries compared to small countries, COI citations of each paper range from 0.156 to 0.883, while the lower $COIC/P_n$ values are similar to the ones in Fig.~\ref{Figure5}c.

\begin{figure*}[htbp]
  \centering
  \includegraphics[width=1\textwidth]{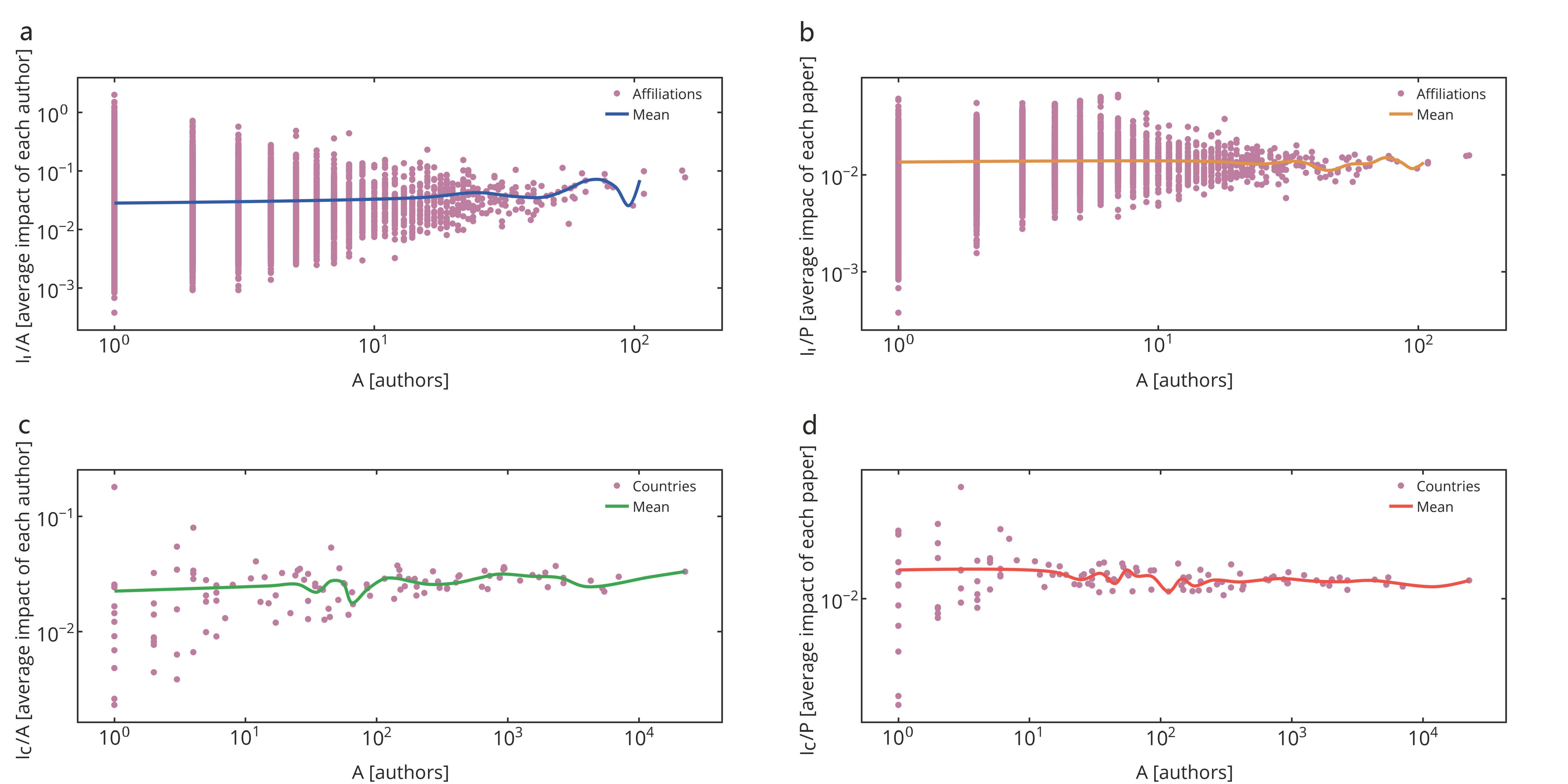}
  \caption{Quantitative patterns of impact for institutions and countries. (Note: (a) Institution size positively correlates with the average impact of institutions on the whole. (b) Institution size has no specific influence on the average publications impact of institution. (c) Country size has positive impact on the average impact in different country size as a whole. (d) Country size has a very small influence on the average publications impact.)}
  \label{Figure6}
\end{figure*}

In order to investigate the impact of average institution and average scholarly papers in different institution sizes and country sizes, several external characteristics are examined. The institution impact ($I_I$) represents the total scores of all authors collected by all papers $P_i$ in this affiliation. The country impact ($I_C$), denotes the total summation of all the institutions collected by all the authors in this country. In Figs.~\ref{Figure6}a and \ref{Figure6}b, the correlation between the institution size $A_m$ and both average institution impact $I_I/A_m$ and average scholarly papers impact $I_I/P_m$ are shown. We find that institution size influences the impact of each author in his/her affiliation, that is, scholars in large institution obtain higher average impact (Fig.~\ref{Figure6}a). We also notice that the institution size has little influence on the average impact of publications in institutions on the whole (Fig.~\ref{Figure6}b). In Figs.~\ref{Figure6}c and \ref{Figure6}d, the correlation between the country size $A_n$ and both country average impact $I_C/A_n$ and average country scholarly papers impact $I_C/P_n$ are demonstrated, which indicates country size has some positive influence on the impact of country, however, the country size has little influence on the average impact of publications in countries on the whole. According to Figs.~\ref{Figure5}c and~\ref{Figure6}c, we observe that if the impact of country is small, its COI citations are small.

\begin{figure*}[htbp]
  \centering
  \includegraphics[width=1\textwidth]{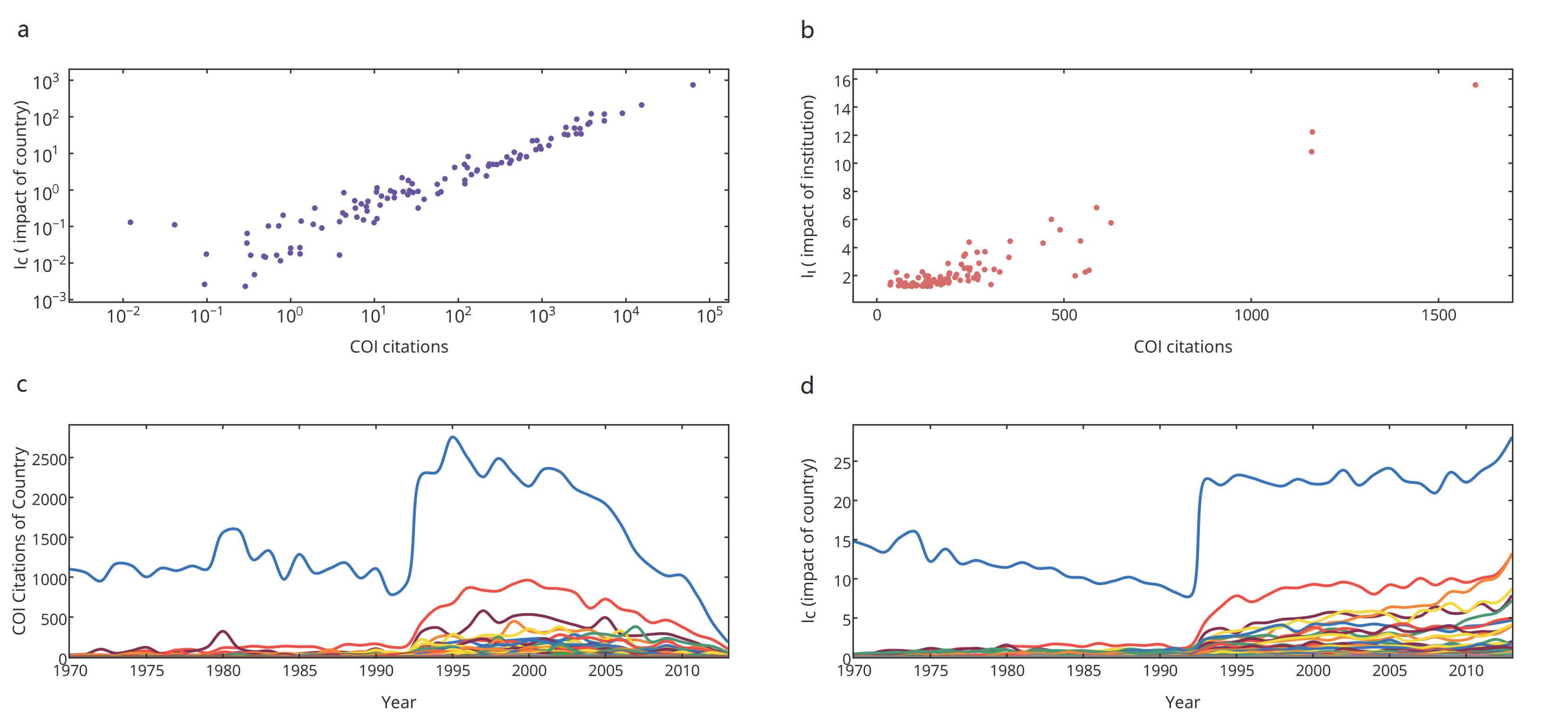}
  \caption{Comparing the impact of institutions and countries against COI citations, and charactering the dynamics of COI citations and the impact of country. (Note: (a) COI citations have a positive correlation with the impact of country $C_i$, each color dot indicates a country. (b) The higher the impact of institution is, the more COI citations it may contain. (c) The yearly variance of COI citations in country-level indicates since 1993, the COI citations have shown the tendency of surge. (d) Striking similarities can be found between the yearly impact of country $C_i(t)$ from 1970 to 2013 and the trend of COI citations of country in Fig.~\ref{Figure7}(c).)}
  \label{Figure7}
\end{figure*}

To investigate the relationship of COI citations and impact of papers evolving at the institution-level and at the country-level, previous studies have been focusing on citation distributions, yet little is concerned about the evolution of COI citations of individual institution and country. The trends of COI citations variations in different institutions and countries are well illustrated by the COI citations history of publications extracted from PRC and PRE subsets of APS corpus (Figs.~\ref{Figure7}a and \ref{Figure7}b). In Fig.~\ref{Figure7}b, we analyze the relationship between the impact of top 100 institutions and their COI citations. The relationship distribution indicates that the higher the impact of institution is, the more COI citations it may contain. For example, the impact and COI citations of the top ranked institution are far beyond the other institutions. Most institutions with lower COI citations have a relative smaller impact than other institutions with higher COI citations. According to Fig.~\ref{Figure7}b, we observe that the top 20 institutions are from American national laboratory and university by the impact of institutions rank. In Fig.~\ref{Figure7}a, we observe that the countries with higher impact likely have more COI citations than the countries with lower impact. By COI citations of each country rank, top 20 countries from high to low in order are: America, Germany, France, Japan, Italy, China, UK, Spain, Australia, Canada, India, Netherlands, Russia, Belgium, Brazil, Israel, Poland, Switzerland, Denmark and Sweden.

According to Fig.~\ref{Figure7}c, we observe that the COI citations of some countries suddenly increase since 1993 year, the trend of high COI citations continues until 2005. After that, COI citations of country present decreased trend. The reason behind is that COI citations of a paper are a cumulative process. The published time of papers becomes shorter, and possible COI citations are relatively smaller, therefore, we observe that around 2013, the COI citations of country become less. Fig.~\ref{Figure7}d characters the dynamic impact of scholarly papers in country-level. We observe that the impact of countries presents a sudden increase since 1993, which is consistent with Fig.~\ref{Figure7}c. What fueled this increase in the impact of country? From the statistics of the yearly summation of national publications, we observe that the number of publications grows rapidly in 1993, and some countries even increase to 10 times, compared to 1992. We believe this shift has close relationship with the "information superhighway" strategy of the United States originated from the Clinton period. In September 1993, shortly after Clinton became the president of the United States, he officially launched the cross-century national information infrastructure project plan. This program not only had a very broad impact on the world, but also created a brilliant future for the United States information economy. In addition, a journal paper published in Nature 2013 also reported that international collaboration increased more than ten-fold since the mid-1990s, which coincides with our experimental results from another aspect~\cite{adams2013collaborations}. Top 20 countries from high to low in order are: America, Germany, France, China, Japan, India, Italy, UK, Spain, Brazil, Russia, Canada, Netherlands, Australia, Israel, Belgium, Poland, Argentina, Korea and Switzerland by impact of country rank. Specially, America is the most prominent country, and its impact is about 3.5 times of Germany. Through such sorting results, we observe that the ranking order of country impact correlates with scientific and technological competitiveness and more details for scientific competitiveness of nations can be found in the reference~\cite{bornmann2014effect}.

\section{Discussion}
Our analysis about the citation relationship in APS dataset by time evolution indicates that regardless of the institution size, the COI relationships in scientific evolution networks are a common phenomenon and COI citations present variable trends in different institutions and countries. Institutions with the highest COI citations are found in USA, followed by Australia, Germany, Israel, Finland and Spain. While USA and Germany belong to G8 countries, Australia and Israel are the countries with high investments in research~\cite{cimini2014scientific}. This finding suggests that leading countries in the world are not insulated from negative citations behaviors. Understanding how the COI relationships affect the appraisal of scholarly papers is of great importance for scientific community. We investigate the criteria of fair assessment of the scientific impact, and the proposed PANDORA technique helps identifying highly influential papers, scholars, and academic institutions.

There are several potential explanations for why COI relationships exist in citation networks. Firstly, the cooperation under the competition mechanism may play a crucial role. Specially, an extremely strong collaboration relationship (super ties) has remarkably positive impact on citations~\cite{petersen2015quantifying}. For example, pursuing collaborative research in a close community tends to be more beneficial than working alone, and negative COI citations could increase at the same time. Secondly, some researchers may cite deliberately each other¡¯s work to boost citation count, hence, the implied academic reputations~\cite{yao2014ranking}, i.e., researchers may cite their friends' newly published papers. Lastly, the same scientific affiliation may also contribute to negative COI citations, i.e., they likely cite their colleagues just because they belong to a scientific team instead of actual research relevance. Although we have stated several possible reasons of COI citations generation, the behavior in real scientific community are more complicated and further investigating in COI citations is required to fairly evaluate the scientific impact.

In summary, we have proposed a COI-based method to give the audiences a deeper insight into the impact of papers, institutions and countries. Our method has several distinct features over conventional approaches: (i) The citing strength is determined based on the COI relationships in citation networks, and the four categories of COI relationships are investigated for adjusting the citation weights. To this end, a co-citation mechanism is used to identify positive COI, negative COI, positive suspected COI and negative suspected COI relationships in citation networks. In our method, a homogeneous and heterogeneous fusion perspective is adopted so that the scores of PageRank, authors, journals, references fit together into the corresponding impact of articles calculation. To quantify the impact of papers, the well-known approach PageRank treats all citation weights as 1, making all citations with equal importance. However, as COI relationships exist, our approach adopts a weighted PageRank to identify different citation relationships to fairly evaluate the impact of papers. (ii) PANDORA performs better than the current methods in terms of evaluating the impact of institutions. We adopt a credit allocation algorithm for fairly distributing the impact of each publication to its authors, which could guarantee accurate appraisal of authors' impact. Then we reasonably assign the authority score of each scholar to his/her affiliation and country. Previous research equally allocated the impact of single paper to its all authors, or distributed them according to the sequence order of authors. Thus some papers are computed twice or more when calculating the total publications of a country~\cite{king2004scientific}. In our method, we take the first author¡¯s first affiliation to resolve these issues. (iii) Our approach could be utilized to improve some established metrics for scientific impact by considering the COI relationships, such as JIF and H-index. Current evaluation methods of citation-based impact, from JIF to H-index, contain anomalous citations inevitably leading to the inaccuracy of evaluation. However, PANDORA can effectively discover such abnormal citations. If the abnormal citations can be eliminated before calculating JIF and H-index, these metrics can reflect true impact more effectively.

Our proposed method has few limitations: (i) Our experiments covered a descent number of research papers (71,287), these papers come from APS dataset and the study is limited to the physics discipline. (ii) Dependence of the data quality: for example, we extracted the affiliation details from APS dataset, thus relied on the mechanism that APS chosen to combine multiple affiliations that are variants of the same institution. (iii) COI relationships are identified through co-author and same institution, whereas other forms of COI relationships might also exist.

Our analysis provides guidance on quantifying the impact of the academic publications, institutions and countries using a metrics-based system. As such, understanding how research papers trajectories shift in affiliation-level and country-level is valuable for career planning and selection of collaborative opportunities. Furthermore, better understanding of the COI relationships in citation networks is also significant for assessing efficiency and productivity of scientific research.

\section*{Acknowledgment}
The authors extend their appreciation to the International Scientific Partnership Program ISPP at King Saud University for funding this research work through ISPP$\#$0078. This work was supported in part by the National Natural Science
Foundation of China under Grant 61502075, China Postdoctoral
Science Foundation Funded Project (2015M580224), and Liaoning Province Doctor Startup Fund (201501166).

\bibliographystyle{IEEEtran}
\bibliography{IEEEabrv}
\vfill
\end{document}